\documentclass[preprint,showpacs,preprintnumbers,floats,floatfix,nofootinbib]{revtex4}

\usepackage{graphicx}
\usepackage{dcolumn}
\usepackage{bm}

\newcommand{\bea}{\begin{eqnarray}}
\newcommand{\eea}{\end{eqnarray}}
\newcommand{\be}{\begin{equation}}
\newcommand{\ee}{\end{equation}}
\newcommand{\p}{{\phi}}
\newcommand{\pd}{\dot{\phi}}
\newcommand{\pdd}{\ddot{\phi}}
\newcommand{\ld}{\dot{\lambda}}
\newcommand{\ldd}{\ddot{\lambda}}
\newcommand{\nd}{\dot{\nu}}
\newcommand{\ndd}{\ddot{\nu}}

\begin{document}


\title{Stabilization of Extra Dimensions at Tree Level}

\author{Scott Watson}
 \email{watson@het.brown.edu}
\author{Robert Brandenberger}%
 \email{rhb@het.brown.edu}
\affiliation{Department of Physics, Brown University, Providence, RI.}

\date{\today}

\begin{abstract}

By considering the effects of string winding and momentum modes on
a time dependent background, we find a method by which six compact
dimensions become stabilized naturally at the self-dual radius
while three dimensions grow large.
\end{abstract}

\pacs{Valid PACS appear here}
\maketitle

\section{Introduction}
String theory continues to be a promising candidate for a quantum
theory of gravity.  However, there are several key challenges
in an attempt to relate the theory to phenomenology.  One such issue is
that string theory predicts a number of extra spatial dimensions.
A standard resolution to this problem is to
assume that six of the dimensions are small enough to escape
experimental detection, which usually means they are taken to be on the
order of the Planck scale.  Given this, we must not only
explain why the six spatial dimensions evolve differently from the other
three, but also why they are frozen at such an extraordinarily small size.

A possible resolution to this problem was suggested in
\cite{bv} (see also \cite{vafa}) and has since been generalized to include
more realistic background geometries and the
inclusion of branes \cite{extended,damien,me}.  The authors of \cite{bv}
argued that, by considering the dynamics of the winding and momentum modes
of the strings, a new symmetry specific of string theory, namely t-duality,
could eliminate the big-bang
singularity and also explain the dimensionality of
space-time.  However, these arguments were only qualitative.

In this paper we want to quantify some of the arguments presented in \cite{bv}.
Specifically we will demonstrate that, working in the regime of
weak string coupling and at tree level in $\alpha^{\prime}$, we find
that the dynamics of string winding and momentum modes lead to a
natural mechanism to stabilize the extra dimensions at the self-dual radius.

In Section 2 we will review briefly the origin of the equations
of motion for strings in time-dependent backgrounds.  These result
from demanding conformal invariance of the world-sheet action
(vanishing $\beta$ functions).  In Section 3 we
include the massive modes of the string as source terms for the {\em stringy}
Einstein equations and demonstrate that the complete set of equations are
invariant under t-duality. In Section 4 we solve the
equations in the presence of string sources and find that
stability of the extra dimensions results naturally for arbitrary
initial conditions.  We conclude with some brief remarks.

\section{Dynamics of Strings in Time-Dependent Backgrounds}

We begin this section by briefly reviewing the
equations of motion for strings in time-dependent backgrounds
(for more details see e.g. \cite{gsw}).
The starting point for strings in a curved space-time is the nonlinear
sigma model whose action is
%
\be
S_{\sigma}=-\frac{1}{4 \pi \alpha^{\prime}}\int d^{2}\sigma
\Bigl( \sqrt{-\gamma} \gamma^{ab}G_{\mu \nu}(X) \;
\partial_{a}X^{\mu}\partial_{b}X^{\nu}+
\epsilon^{ab} B_{\mu \nu}(X)\;
\partial_{a}X^{\mu}\partial_{b}X^{\nu}+
\alpha^{\prime} \sqrt{\gamma} \phi(X)  R^{(2)} \Bigr), \ee
%
where $\gamma^{ab}$ is the world-sheet metric,
$(2 \pi \alpha^{\prime})$ is the inverse string tension,
$G_{\mu \nu}$ is the background space-time metric,
$B_{\mu \nu}$ is the background antisymmetric tensor, and $\phi$ is the
background dilaton which is coupled to the world-sheet Ricci scalar
$R^{(2)}$.  The string coupling is given in terms of the dilaton
by $g_{s}=e^{2\phi}$.

We obtain the equations of motion by demanding conformal invariance
of the world-sheet action.  This is equivalent to demanding
that the trace of the world-sheet stress
tensor given by
\be
2\pi T^{a}_{a}= \beta^{G}_{\mu \nu}\sqrt{\gamma}\gamma^{ab}\partial_{a}X^{\mu}\partial_{b}X^{\nu}
+\beta^{B}_{\mu \nu}\epsilon^{ab}\partial_{a}X^{\mu}\partial_{b}X^{\nu}+\beta^{\phi}\sqrt{\gamma}R^{(2)},
\ee
vanish, where the $\beta$ functions are
\bea
\label{betaeqns}
& & \beta^{G}_{\mu \nu}= \alpha^{\prime}\Bigl( R_{\mu \nu} +2 \nabla_{\mu} \nabla_{\nu}\phi
-\frac{1}{4} H_{\mu \kappa \sigma} H_{\nu}^{\; \kappa
\sigma}\Bigr)
+ {\cal O}(\alpha^{\prime 2}),\nonumber \\
& & \beta^{B}_{\mu \nu}=\alpha^{\prime}\Bigl( \nabla^{\kappa}H_{\kappa \mu \nu}
- 2\nabla^{\kappa}\phi H_{\kappa \mu \nu }\Bigr)+ {\cal O}(\alpha^{\prime 2}),\nonumber \\
& & \beta^{\phi}= \alpha^{\prime}\Bigl( \frac{D-26}{3 \alpha^{\prime}}-4\nabla_{\kappa}\nabla^{\kappa}\phi
+4\nabla_{\kappa}\phi \nabla^{\kappa} \phi-R +\frac{1}{12}H_{\kappa \mu \nu} H^{\kappa \mu
\nu}\Bigr)
+ {\cal O}(\alpha^{\prime 2}),
\eea
with $H=dB$ denoting the field strength associated with the field
$B_{\mu \nu}$. Keeping terms to lowest order in $\alpha^{\prime}$,
these equations of motion
can alternatively be derived from a low energy effective action
formulated in the target space or {\em bulk},
\be
\label{bulkaction}
S_{bulk}=\frac{1}{4\pi \alpha^{\prime}}\int d^{D}x \sqrt{-G} e^{-2 \phi}\Bigl( R+4(\nabla \phi)^{2}-\frac{1}{12}H^{2}
\Bigr) \, ,
\ee
where $R$ is the Ricci scalar and $G$ is the determinant of the metric.
Thus far we have restricted ourselves to the bosonic
string, however this action remains valid for the case of the
supersymmetric string.  For example,
with $D=10$ Eq. (\ref{bulkaction}) becomes the low energy effective action of
type II-A superstring theory.

We now proceed by demanding that the $\beta$ functions vanish. Assuming
that we are in the critical dimension and that there are no fluxes (i.e. $D=26 \; or \; 10$ and $B_{\mu
\nu}=0$) we find that (\ref{betaeqns}) becomes,
\bea
\label{eoma}
& & R_{\mu \nu} +2 \nabla_{\mu} \nabla_{\nu}\phi
=0, \nonumber \\
& & R+4\nabla_{\kappa}\nabla^{\kappa}\phi
-4\nabla_{\kappa}\phi \nabla^{\kappa} \phi=0.
\eea

Next we wish to include {\em stringy} sources into
the modified Einstein equations (\ref{eoma}).  This can be done by adding a
matter term to the bulk action (\ref{bulkaction}) as was done in
\cite{vafa}.  Here we will take a different approach by including
the sources at the level of the equations of motion in the form of
the stress energy tensor.  We expect the equations of motion to
generalize in the presence of string sources to
\bea
\label{eomb}
& & R_{\mu}^{\; \; \nu} +2 \nabla_{\mu} \nabla^{\nu}\phi
=8 \pi M^{-2}_{p} \; e^{2\phi} T_{\mu}^{\; \; \nu}, \nonumber \\
& & R+4\nabla_{\kappa}\nabla^{\kappa}\phi
-4\nabla_{\kappa}\phi \nabla^{\kappa} \phi=0.
\eea
We will assume that the string sources take the form of a perfect
fluid,
\be
T_{\mu}^{\: \:
\nu}=diag(\rho,-p_{1},-p_{2},-p_{3},-p_{4},-p_{5},-p_{6},-p_{7},-p_{8},-p_{9}),
\ee
where $\rho$ is the energy density of the strings and $p_i$ is the
pressure density in the i'th direction.

We conclude this section by consider the equations of
motion (\ref{eomb}) under the assumption of a homogeneous metric
of the form
\be
ds^{2}=dt^{2}-e^{2 \lambda} d\vec{x}^{2}-e^{2 \nu} d\vec{y}^{2},
\ee
where $(t,\vec{x})$ are the coordinates of $3+1$ space-time and
$\vec{y}$ are the coordinates
of the other six dimensions.  The scale factors $a(t)$ and $b(t)$
are defined by
$\lambda \equiv \ln(a(t))$ and $\nu \equiv \ln(b(t))$.

Given this ansatz for the metric and assuming the string sources to be
a perfect fluid, the equations of motion (\ref{eomb}) become
\bea
\label{eomc1}
& & -3\ldd-3\ld^{2}-6\ndd-6\nd^{2}+2\pdd=\frac{1}{2}e^{2\p}\rho,\\
\label{eomc2}
& & \ldd+3\ld^{2}+6\ld\nd-2\ld\pd=\frac{1}{2}e^{2\p}p_{\lambda},\\
\label{eomc3}
& & \ndd+6\nd^{2}+3\ld\nd-2\nd\pd=\frac{1}{2}e^{2\p}p_{\nu},\\
& &
\label{eomc4}
3\ldd+6\ld^{2}+6\ndd+21\nd^2+18\ld\nd+4\pd^2-4\pdd-12\ld\pd-24\nd\pd=0,
\eea where $p_{\lambda}$ and $p_{\nu}$ are the pressures in the
respective dimensions and we work in Planck units with
$16 \pi M_{p}^{-2} = e^{2 \phi}$.

We note that setting the dilaton to a constant takes our
equations to the expected Friedmann-Robertson-Walker (FRW) equations with
the constraint
$R=0$.  Explicitly, if we restrict to the $3+1$ dimensional case ($\nu=0$)
we find,
\bea
& & H^{2}\equiv \ld^{2}=\frac{8 \pi}{3 M_{p}^{2}} \rho\\
& & \ldd+\ld^{2}= -\frac{4 \pi}{ 3 M_{p}^{2}}(\rho+3 p) \\
& & R=\ldd+2\ld^{2}=0.
\eea
The last condition, $R=0$, implies $T^{\; \;\mu}_{\mu}=0$, which
tells us that as the Einstein theory becomes the effective theory
the evolution must start in a {\em radiation-like} phase with
$\rho-3p=0$.  This is no surprise since the equations were
obtained by demanding conformal invariance.  Moreover, this ties
the picture together nicely since we expect the {\em stringy}
effects in cosmology to eventually settle into the radiation
dominated phase of the standard cosmological model.

\section{String Sources and T-duality}

In \cite{bv} it was argued that by considering
the dynamics of closed strings on a compact background geometry one
could not only produce a nonsingular cosmology, but also provide an
explanation for the dimensionality of space-time.  The analysis of
\cite{bv} was heuristic but lacked rigorous quantitative calculations.
Here we want to address some of the issues in a more quantitative manor
(for works addressing other issues more rigorously see
\cite{vafa,extended,damien,me}). In particular, we will demonstrate that
a mechanism to stabilize the extra dimensions can result.  Note that a similar investigation
of the role of massive string states on the evolution of small
and large dimensions was recently published in \cite{Borunda}. Our
results agree with those of \cite{Borunda}, although the emphasis
on the stabilization mechanism is new here.

Closed string theories on a compact geometry have their mass spectrum
altered in two ways.  First, because the center of mass momentum
must be periodic in the compact directions we get its quantization
analogous to the Kaluza-Klein case.   In addition to
these {\em momentum} modes there are additional degrees of freedom
associated with the possible winding of the strings.
These {(anti-)}{\em winding} modes wrap the compact dimensions in a
(counter-)clockwise direction.
Associated with the winding is a topologically conserved charge known as the
winding number.  This quantity is (negative) positive for (anti-)
winding modes and
is conserved so winding modes can only be created and destroyed in
pairs.  When a winding mode intersects with an anti-winding mode this
results in a closed unwound string with winding number
zero.  The total mass spectrum of the string also includes the oscillatory
modes which give rise to the particle spectrum.
The spectrum is found by demanding all states to be on-shell and in the
case of one compact dimension takes the form \cite{Polchinski}
\be
\label{massspectrum}
M^2=\frac{n^2}{R^2}+w^2 R^2+2(N+\tilde{N}-2),
\ee
where we have chosen units where $\alpha^{\prime}=1$.  The integers
$n$ and $w$ denote the
Kaluza-Klein level and the winding numbers of the string,
respectively. $N$ and $\tilde{N}$ are the left and right
oscillators of the string that give rise to the particle spectrum and
$R$ is the radius of the compact dimension.
It was shown in \cite{Kripfganz} that the oscillator terms are
exponentially suppressed and rendered
unimportant for determining the overall evolution of the background.
Thus, we will focus on the zero modes of the mass spectrum in the
rest of this paper.

The important result that can immediately be seen from
(\ref{massspectrum}) is that the spectrum remains unchanged if we send
$R\rightarrow 1/R$ and $n \leftrightarrow w$.  This property is
know as t-duality.  It turns out that t-duality is not just a
property of the strings, but also of the cosmological background
\cite{Polchinski_tdual}. To see this, we note that the role
of the radius is played by the scale factors $a(t)$ and $b(t)$.
We then find that the cosmological equations (\ref{eomc1})-(\ref{eomc4})
are invariant under the duality transformation,
\be
\lambda(t) \rightarrow -\lambda(t), \;\;\;\;
\nu(t) \rightarrow -\nu(t), \;\;\;\;
\p(t) \rightarrow \p(t)-3\lambda(t)-6\nu(t).
\ee
Shifting the dilaton by the volume factor is required because this
is a dynamical (time-dependent) duality.

Now that we have observed that our equations and string sources
are duality invariant let us proceed by explicitly constructing
energy and pressure terms consisting of the zero modes of the
string. From (\ref{massspectrum}) we find that the zero mode
energy and pressure \footnote{This pressure is related to the
previous by $P_{i}=p_{i}V$ where $V$ is the volume.} of the string
gas in $D-1$ compact dimensions can be written as, \bea
\label{energypressure} & & E=3\mu N^{(3)}e^{\lambda}+3\mu
M^{(3)}e^{-\lambda}+6\mu N^{(6)}e^{\nu}+6\mu M^{(6)}e^{-\nu},
\nonumber \\ & & P_{\lambda}=-\mu N^{(3)}e^{\lambda}+\mu
M^{(3)}e^{-\lambda},\nonumber \\ & & P_{\nu}=-\mu
N^{(6)}e^{\nu}+\mu M^{(6)}e^{-\nu}, \eea where $\mu$ is the
chemical potential (mass per unit length of the string), $E$ is
the energy, and $P_{i}$, $N^{i}$, $M^{i}$ are the pressure, number
of winding and momentum modes in the {\em i}th direction,
respectively. By substituting these terms into
(\ref{eomc1})-(\ref{eomc4}) we find the equations describing the
evolution of our background in terms of string sources, \bea
\label{eoma1} & &
-\ldd-\ld^{2}-2\ndd-2\nd^{2}+\frac{2}{3}\pdd=\frac{\mu}{2}e^{2\p-3\lambda-6\nu}
\Bigl(  N^{(3)}e^{\lambda}+M^{(3)}e^{-\lambda}+2 N^{(6)}e^{\nu}+2
M^{(6)}e^{-\nu}\Bigr)  \\ \label{eoma2} & &
\ldd+3\ld^{2}+6\ld\nd-2\ld\pd=\frac{\mu}{2}e^{2\p-3\lambda-6\nu}
\Bigl(-N^{(3)}e^{\lambda}+M^{(3)}e^{-\lambda}\Bigr) \\
\label{eoma3} & &
\ndd+6\nd^{2}+3\ld\nd-2\nd\pd=\frac{\mu}{2}e^{2\p-3\lambda-6\nu}
\Bigl(-N^{(6)}e^{\nu}+M^{(6)}e^{-\nu}\Bigl)\\ & & \label{eoma4}
3\ldd+6\ld^{2}+6\ndd+21\nd^2+18\ld\nd+4\pd^2-4\pdd-12\ld\pd-24\nd\pd=0.
\eea These equations give us a quantitative way to address the
issues first discussed in \cite{bv}, where it was argued that the
winding modes of the strings become more massive as the universe
expands, thus preventing the universe from expanding. We can see
this through the above equations since the winding modes
contribute a negative pressure term and thus a negative effective
potential in the equations of motion for $\lambda$ and $\mu$
\cite{vafa}. Thus, the negative pressure of the winding modes does
{\em NOT} imply an accelerating phase for the background as it
would if the background were described by pure General Relativity.

When considering the initial state to consist of a gas of string winding modes,
the dimensionality and isotropy of space-time can be explained as a natural
consequence of the dynamics \cite{bv,damien,me}.
The strings are initially taken to be in thermal equilibrium and pairs of wound
strings are created and annihilated allowing expansion to persist.
As the expansion continues the winding modes fall out of
equilibrium and the negative pressure of the remaining modes will halt the
expansion.
This leads to a period of loitering at which time strings in three
of the dimensions can find each other and annihilate into
loops \cite{damien}.  This leaves three dimensions filled with a gas of string
loops with an equation of state resembling ordinary radiation,
whereas the other six dimensions remain compact.  Therefore, the
dimensionality of space-time results from decompactification of
three dimensions, since this is the maximum number of
dimensions that strings are able to find each other to intersect.

There are several points of concern with the above argument. As
the three dimensions grow large the strings that have not yet
annihilated and the strings in the six small dimensions could play
an important role in the dynamics.  Considering the effect that
these inhomogeneities have on the geometry and stability of the
model is important for the success of the model and is currently
being examined \cite{watson}. Another important issue is the
stability of the internal dimensions.  In the above argument it
was assumed that the dimensions are trying to expand, but we must
also consider the case of contraction.  This is where the momentum
modes play an important role in the dynamics.  As mentioned above,
the momentum modes are dual to the winding modes and result in the
opposite dynamical behavior. That is, as the universe collapses
these modes become heavy and it becomes energetically favored to
re-expand.  In this way the momentum modes prevent collapse to a
singularity by contributing an increasing positive pressure to
drive the evolution towards expansion.  Thus, the resulting
cosmology is non-singular.

Considering both the winding and momentum modes suggests that the
natural size of the universe should be at the self-dual radius
where the total energy is minimized. At this radius the negative
pressure of the winding modes is exactly canceled by the positive
pressure of the momentum modes. Thus, in the context of string
theory it is natural to expect the evolution of our universe to
begin at the self-dual radius, which is unity in string units. In
fact, this radius is a very special radius in string theory and
represents a point of enhanced symmetry for the gauge groups
associated with the internal dimensions and the strings (c.f.
\cite{johnson}).

An important point that we have not yet discussed is the role of
the dilaton. Recall that the string coupling is given by
$g_{s}=e^{2 \phi}$, where $\phi$ is the dilaton.  In order for the
equations of motion (\ref{eoma1})-(\ref{eoma4}) to remain valid we
must restrict the phase space to the region of small string
coupling ($g_{s} <<1$). Moreover, we must choose initial
conditions that do not result in a rapidly growing coupling.  In
our analysis we simply enforce this as an energetically favored
constraint. Moreover, a more complete analysis would consider a
potential for the dilaton.  This would allow us to take our
considerations out of the {\em stringy} regime and into the
classical FRW radiation dominated universe.  The potential of the
dilaton should be provided by the correct model of supersymmetry
breaking (this remains one of the outstanding challenges for
string theory).

\section{Stabilization}

It was shown in \cite{damien} and \cite{me} that
the equations of motion (\ref{eoma1})-(\ref{eoma4}) lead to a period of
cosmological loitering, allowing winding modes to annihilate in three
dimensions and resulting in those three dimensions expanding to a large
size. However, in these approaches it was assumed that the effect
of the six small dimensions could be ignored. Here we improve
on this analysis and include the evolution of the small dimensions as well
as the gravitational coupling between the large and the small dimensions.

We are interested in Eqs. (\ref{eoma1})-(\ref{eoma4}) in the case
when three of the spatial dimensions are taken to be large,
expanding, and filled with momentum modes only, i.e. $N^{(3)}=0$,
since the winding modes have all annihilated into string loops.
The other six dimensions are taken to start at the self-dual
radius, i.e. $b = 1$ and ${\dot b} = 0$ in string units.  This
implies $\nu=\dot{\nu}=\ddot{\nu}=0$, and thus it follows from
(\ref{eoma3})
\be
\ndd+6\nd^{2}+3\ld\nd-2\nd\pd=0=\frac{\mu}{2}e^{2\p-3\lambda-6\nu}
\Bigl(-N^{(6)}e^{\nu}+M^{(6)}e^{-\nu}\Bigl),
\ee
that the pressure must vanish, which is only possible if
\be \label{constraint}
N^{(6)} \, = \, M^{(6)} \, .
\ee
This is an expected result, since at the self-dual point the momentum modes
and winding modes should be equivalent.

Given the constraint (\ref{constraint}) (and making use of the
notation $l(t)=\ld$, $q(t)=\nd$, and $f(t)=\pd$) we can
rewrite the system (\ref{eoma1})-(\ref{eoma4}) as the following system
of first order differential equations,
\bea
\label{constraintequation}
& & 7 l^{2}+35 q^2+42 lq+\frac{4}{3}f^2-10lf-20qf \nonumber \\
& & \; \; \; \; \; \; \; \;=\mu e^{2\phi-3\lambda-6\nu} \Bigl( 5 M^{(3)} e^{-\lambda}+8 N^{(6)} \cosh(\nu)-3N^{(6)}
\sinh(\nu) \Bigr)\\
\label{eq1}
& & \dot{f}= -\frac{15}{4}q^2-\frac{9}{2}lq-3qf-\frac{3}{4}l^2-\frac{3}{2}lf+f^2 \nonumber \\
& & \; \; \; \; \; \; \; \;+\frac{3\mu}{4} e^{2 \phi-3\lambda-6\nu}\Bigl( \frac{1}{2}
M^{(3)}e^{-\lambda}- 2N^{(6)} \sinh(\nu) \Bigr) \\
\label{eq2}
& & \dot{l} = -3l^2-6lq+2lf+\frac{\mu}{2}M^{(3)} e^{2\phi-4 \lambda - 6
\nu} \\
\label{eq3}
& & \dot{q} = -6q^2-3lq+2qf- \mu N^{(6)} e^{2 \phi-6 \nu - 3 \lambda}
\sinh(\nu).
\eea
We take (\ref{constraintequation}) as a
constraint on the initial data and then solve the remaining system
numerically.

\begin{figure}[!]
\includegraphics[totalheight=3 in,keepaspectratio]{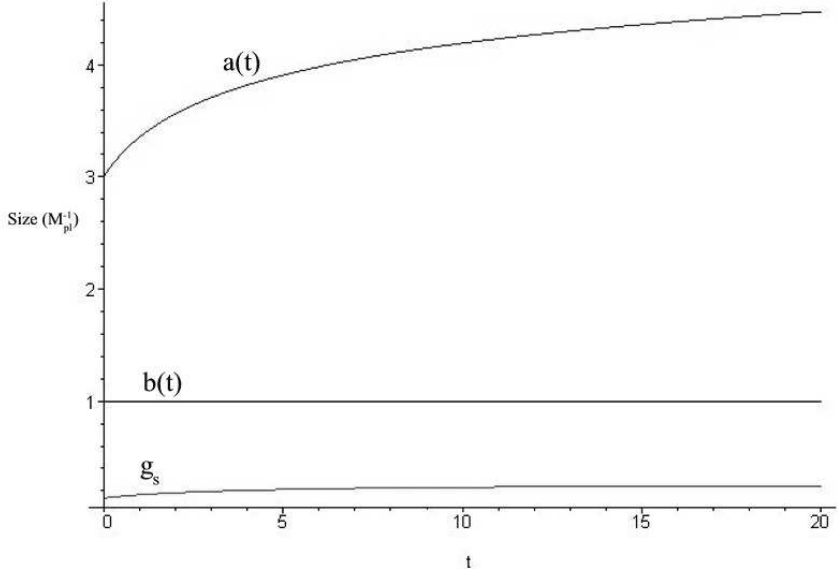}
\caption{Here we take the six small dimensions to be at the
self-dual radius and with vanishing expansion rate initially,
$b_{0}=1$ and $\dot{\nu}_{0}=0$.  We find that the
dimensions remain stable at the self-dual point regardless of the
behavior of the dilaton and of the three large dimensions.  These are
shown in the figure for the initial values $\lambda_{0}=3$,
$\ld_{0}=0.5$, and $\phi_{0}=-3$. However, this result holds for
generic initial values as long as we respect the weak coupling
limit (i.e. $g_{s}<<1$).  Note that we are using Planck units.
} \label{fig1}
\end{figure}

\begin{figure}[!]
\includegraphics[totalheight=3 in,keepaspectratio]{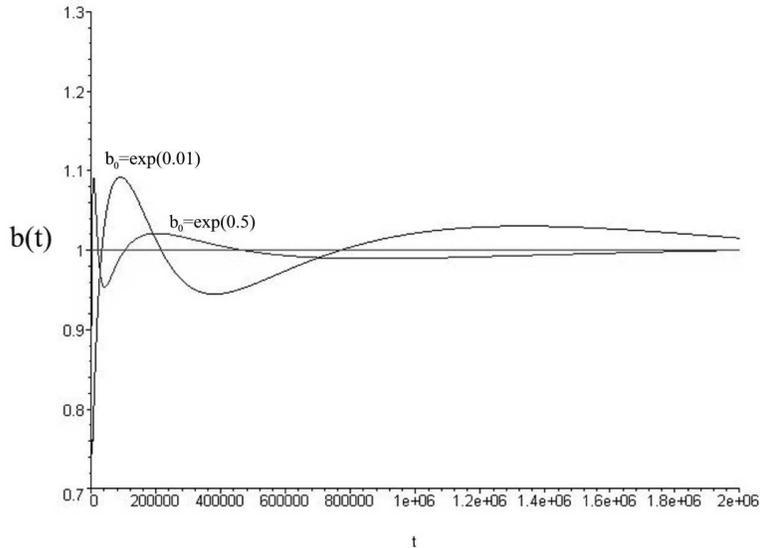}
\caption{We consider initial values of $b(t)$ away from the
self-dual radius and find oscillations about the self-dual radius
which are damped by the dilaton and by the evolution of the large
dimensions. This is again a generic result.  Here we present the
evolution for the following initial data: {$ \dot{\lambda}_{0} =
.500, \phi_{0} = -3.00, \pd_{0} = .380, \lambda_{0} = 3.00,
\nd_{0} = 0, \nu_{0} = 0$}\\{$\ld_{0} = .500, \phi_{0} = -3.00,
\pd_{0} = -.020, \lambda_{0} = 3.00, \nd_{0} = -.100, \nu_{0} =
.010$}\\ {$\ld_{0} = .500, \phi_{0} = -3.00, \pd_{0} = -.008,
\lambda_{0} = 3.00, \nd_{0} = -.100, \nu_{0} = .500$}}
\label{fig2}
\end{figure}

We first consider the six small dimensions to be initially static at the
self-dual radius.  We find that they remain fixed and stable at the self-dual
radius for all subsequent times.  This result is robust
and independent of the behavior
of the three large spatial dimensions and the dilaton.  In Fig.
(\ref{fig1}) we plot (in Planck units) the behavior of the two scale factors
and of the string coupling as a function of time.  As can be seen from the
figure, the string coupling remains small for all times
guaranteeing that the weak coupling regime is valid.

We now relax our assumption that the small dimensions begin at the
self-dual point (maintaining, however, the constraint (\ref{constraint})).
In this way we can examine whether the winding
and momentum modes do indeed drive the system towards the self-dual
point. We begin by introducing an initial radius slightly larger
than the self-dual radius and consider the evolution with a nonzero expansion
rate. At the beginning of the evolution, $b(t)$ oscillates
around the self-dual radius with decreasing amplitude as can be seen in
Fig. (\ref{fig2}). The dilaton and the three other dimensions play an
important role by damping the oscillations. It is important to note that the
damping is not dependent on the dilaton alone, as can be seen from
the exponential term in (\ref{eq1})-(\ref{eq3}) \footnote{Recall
that we are interested in the region of phase space where
$g_{s}=e^{2\phi}=e^{-2|\phi|}$.}. The growth of the three large
dimensions is also important for the stabilization process. As the three
large dimensions expand, this damps the oscillation of the
internal scale factor of the six dimensions to the self-dual
radius.

We find that as we increase the initial value of the internal dimensions that
the subsequent motion is damped but with a decreasing frequency.
In Fig. (\ref{fig2}) we consider three cases differing in
the initial values for the size of the small dimensions and their
expansion rate.  We find that if we displace the scale factor of the small
dimensions by an arbitrary amount that the damping will suffice
to drive the evolution of $b(t)$ to
the self-dual point.  In this way we see that the inclusion of the momentum and
winding modes offers a mechanism to stabilize the extra dimensions
at the self-dual radius.

\section{Conclusion}

By considering the effects of string winding and momentum modes on
a time-dependent background, we have shown that stabilization of
extra dimensions results for reasonable initial conditions.
Furthermore, we have shown that the stabilization radius is the
expected self-dual point where the symmetries of the theory are
enhanced.  We remind the reader that we have restricted our
analysis to the weak coupling region of phase space where
$g_{s}<<1$ and worked to lowest order in $\alpha^{\prime}$.

This result is encouraging, since it agrees well with the
predictions of \cite{bv}, which were based on assuming
t-dual matter sources and their plausible effects on the background geometry.
It would be interesting to test the stability of our model under
corrections of higher order in $\alpha^{\prime}$, and also in the
presence of inhomogeneities.  Inhomogeneity is also an important
consideration for the evolution of the three large dimensions and
will be the subject of future work \cite{watson}.

Lastly, we stress again that our
model remains incomplete without a better understanding of
the dilaton.  A successful method to generate a potential for the
dilaton and carrying us from the string theory
regime to the late time phase when classical general relativity
applies is still needed.
Our knowledge of the non-perturbative aspects of string theory
continues to grow, and this may help resolve this problem
and yield a better picture of string theory phenomenology.

\begin{acknowledgments}

RB was supported in part by the U.S. Department of Energy under
Contract DE-FG02-91ER40688, TASK A. SW was supported in part by
the NASA Graduate Student Research Program. SW would also like to
thank S. Cremonini and H. de Vega for useful discussions.

\end{acknowledgments}


\begin{thebibliography}{99}
\bibitem{bv}
R.~H.~Brandenberger and C.~Vafa,
``Superstrings In The Early Universe,''
Nucl.\ Phys.\ B {\bf 316}, 391 (1989).

\bibitem{vafa}
A.~A.~Tseytlin and C.~Vafa,
``Elements of string cosmology,''
Nucl.\ Phys.\ B {\bf 372}, 443 (1992) [arXiv:hep-th/9109048].

\bibitem{extended}
S.~Alexander, R.~H.~Brandenberger and D.~Easson,
``Brane gases in the early Universe,''
Phys.\ Rev.\ D {\bf 62}, 103509 (2000) [arXiv:hep-th/0005212];
R.~Easther, B.~R.~Greene, M.~G.~Jackson and D.~Kabat,
``Brane gas cosmology in M-theory: Late time behavior,''
Phys.\ Rev.\ D {\bf 67}, 123501 (2003) [arXiv:hep-th/0211124];
T.~Boehm and R.~Brandenberger,
``On T-duality in brane gas cosmology,''
arXiv:hep-th/0208188;
D.~A.~Easson,
``Brane gas cosmology and loitering,''
arXiv:hep-th/0111055;
R.~Easther, B.~R.~Greene and
M.~G.~Jackson,
``Cosmological string gas on orbifolds,''
Phys.\ Rev.\ D {\bf 66}, 023502 (2002) [arXiv:hep-th/0204099];
A.~Campos,
``Late-time dynamics of brane gas cosmology,''
arXiv:hep-th/0304216;  A.~Kaya,
``On winding branes and cosmological evolution of extra dimensions in  string theory,''
arXiv:hep-th/0302118;
A.~Kaya and T.~Rador,
``Wrapped branes and compact extra dimensions in cosmology,''
arXiv:hep-th/0301031;
A.~Kaya,
``On winding branes and cosmological evolution of extra dimensions in  string theory,''
arXiv:hep-th/0302118.;
S.~H.~Alexander,
``Brane gas cosmology, M-theory and little string theory,''
arXiv:hep-th/0212151.

\bibitem{Kripfganz}
J.~Kripfganz and H.~Perlt,
Class.\ Quant.\ Grav.\  {\bf 5}, 453 (1988).

\bibitem{damien}
R.~Brandenberger, D.~A.~Easson and D.~Kimberly,
``Loitering phase in brane gas cosmology,''
Nucl.\ Phys.\ B {\bf 623}, 421 (2002) [arXiv:hep-th/0109165].

\bibitem{me}
S.~Watson and R.~H.~Brandenberger,
``Isotropization in brane gas cosmology,''
Phys.\ Rev.\ D {\bf 67}, 043510 (2003) [arXiv:hep-th/0207168].

\bibitem{Borunda}
B.~A.~Bassett, M.~Borunda, M.~Serone and S.~Tsujikawa,
``Aspects of string-gas cosmology at finite temperature,''
Phys.\ Rev.\ D {\bf 67}, 123506 (2003)
[arXiv:hep-th/0301180].

\bibitem{gsw}
M.~B.~Green, J.~H.~Schwarz and E.~Witten, {\it
``Superstring Theory''}, Vols. 1 and 2; {Cambridge Univ. Pr.
(1987) (UK)  (Cambridge Monographs on Mathematical Physics)}.

\bibitem{Polchinski}
J. Polchinski, {\it ``String Theory''}, Vols. 1 and
2; {Cambridge Univ. Pr. (1998) (UK) (Cambridge Monographs
on Mathematical Physics)}.

\bibitem{Polchinski_tdual}
E.~Smith and J.~Polchinski,
``Duality survives time dependence,''
Phys.\ Lett.\ B {\bf 263}, 59 (1991).


\bibitem{watson}
S.~Watson and R.~H.~Brandenberger, in preparation.

\bibitem{johnson}
C.~V.~Johnson,
``D-brane primer,'' arXiv:hep-th/0007170.





\end{thebibliography}

\end{document}